\documentclass[runningheads]{llncs}
\usepackage[T1]{fontenc}
\usepackage[nomargin,inline,marginclue,draft]{fixme}
\fxsetup{status=draft}
\fxsetup{theme=color, mode=multiuser} \FXRegisterAuthor{me}{ame}{\color{red}Me}
\usepackage{tabularx}
\usepackage{booktabs}
\usepackage{pdflscape} 
\usepackage{amssymb} 
\usepackage{pifont}
\usepackage{afterpage}
\newcommand{\cmark}{\ding{51}}%
\newcommand{\xmark}{\ding{55}}%

\usepackage{multirow}
\usepackage{graphicx}
\usepackage[table]{xcolor}

\usepackage{float}

\usepackage{etoolbox}
\apptocmd{\sloppy}{\hbadness 10000\relax}{}{}

\usepackage{graphicx}

\usepackage{hyperref}
\usepackage{cleveref}

\usepackage[numbers]{natbib}

\usepackage[nolist]{acronym}
\begin{acronym}
\acro{AURORA}{A Multicenter Analysis of Stereotactic Radiotherapy to the
Resection Cavity of Brain Metastases}
\acro{BL}{BrainLes}
\acro{BLP}{BrainLes preprocessing}
\acro{BraTS}{Brain Tumor Segmentation Challenge}
\acro{BTK}{BraTS Toolkit}
\acro{CNN}{Convolutional Neural Network}
\acro{DL}{Deep Learning}
\acro{DSC}{Dice similarity coefficient}
\acro{NSD}{normalized surface distance}
\acro{DQE}{Deep Quality Estimation}
\acro{FL}{Federated Learning}
\acro{GAN}{Generative Adversarial Network}
\acro{IoU}{Intersection over Union}
\acro{ML}{Machine Learning}
\acro{MRI}{Magnetic Resonance Imaging}
\acro{PGT}{Peak Ground Truth}
\acro{RWMP}{Real World Model Performance}
\acro{NN}{Neural Network}
\acro{mpMRI}{multi-parametric MRI}
\acro{T1n}{native T1-weighted}
\acro{T1c}{contrast enhanced T1-weighted}
\acro{T2w}{T2-weighted}
\acro{FLA}{T2-weighted fluid-attenuated inversion recovery}
\acro{ISLES}{Ischemic Stroke Lesion Segmentation}
\acro{MS}{Multiple Sclerotic}
\acro{AIMS-TBI}{automated identification of moderate-severe traumatic brain injury lesions}
\acro{HIE}{Hypoxic Ischemic Encephalopathy}
\acro{CaPTk}{Cancer Imaging Phenomics Toolkit}
\acro{HD-GLIO}{HD-GLIO brain tumor segmentation tool}
\acro{DeepMedic}{Efficient Multi-Scale 3D Convolutional Neural Network for Medical Image Segmentation}
\acro{Panoptica}{Instance-wise Evaluation of Semantic Model
Predictions}
\acro{PQ}{Panoptic Quality}
\acro{ASSD}{Average Symmetric Surface Distance}
\acro{HD}{Hausdorff distance}
\acro{clDSC}{Centerline Dice}
\acro{SQ}{Segmentation Quality}
\acro{RQ}{Recognition Quality}
\acro{QC}{quality control}
\acro{ROI}{regions of interest}
\acro{CSF}{cerebrospinal fluid}
\acro{PBTs}{Pediatric brain tumors}
\acro{T2H}{T2-hyperintense region}
\acro{ET}{Enhancing tumor}
\acro{ED}{peritumoral edema}
\acro{SNFH}{surrounding non-enhancing FLAIR hyperintensity }
\acro{NETC}{non-enhancing tumor core}
\acro{CC}{Cystic component}
\acro{RC}{resection cavity}
\acro{WT}{Whole tumor }
\acro{RT}{radiation therapy}
\acro{LGG}{low grade glioma}
\acro{HGG}{high grade glioma}
\acro{MICCAI}{International Conference on Medical Image Computing and Computer-Assisted Intervention}
\acro{GLI}{glioma}
\acro{MEN}{meningioma}
\acro{METS}{metastasis}
\acro{PED}{pediatric}
\acro{SSA}{sub-Saharan African}
\acro{GoAT}{Generalizability across tumors }
\acro{GTV}{Gross Tumor Volume}
\acro{DICOM}{Digital Imaging and Communications in Medicine}
\acro{NIfTI}{Neuroimaging Informatics Technology Initiative}
\acro{PACS}{Picture archiving and communication system}
\acro{API}{Application Programming Interface}
\acro{GUI}{graphical user interface}
\end{acronym}

\begin{document}
\setlength\emergencystretch{1.5em}
%
\newcommand*\samethanks[1][\value{footnote}]{\footnotemark[#1]}

\newcommand{\eac}[1]{\emph{\ac{#1}}}
\newcommand{\eacp}[1]{\emph{\acp{#1}}}
\newcommand{\eacf}[1]{\emph{\acf{#1}}}
\newcommand{\eacfp}[1]{\emph{\acfp{#1}}}

\newcolumntype{C}{>{\centering\arraybackslash}X}
\newcolumntype{M}[1]{>{\centering\arraybackslash}m{#1}}
\newcolumntype{L}[1]{>{\arraybackslash}m{#1}}

\newcommand{\gmidrule}[0]{\arrayrulecolor{black!30}\midrule\arrayrulecolor{black}}
\newcommand{\gcmidrule}[1]{\arrayrulecolor{black!30}\cmidrule{2-#1}\arrayrulecolor{black}}
%
\title{\emph{BraTS orchestrator}: Democratizing and Disseminating state-of-the-art brain tumor image analysis}

\titlerunning{\emph{BraTS Orchestrator}}
%
%
\author{Florian Kofler\inst{1,2,3,4}
\and
Marcel Rosier \inst{1,2,8}
\and
Mehdi Astaraki \inst{12,13}
\and
Ujjwal Baid
\inst{14}
\and
Hendrik Möller \inst{4, 5} 
\and
Josef A. Buchner \inst{6}
\and
Felix Steinbauer \inst{11}
Eva Oswald \inst{2,7}
\and
Ezequiel de la Rosa \inst{1}
\and
Ivan Ezhov \inst{4,8}
\and
Constantin von See \inst{9}
\and
Jan Kirschke \inst{3}
\and
Anton Schmick \inst{10}
\and
Sarthak Pati 
\inst{14,18}
\and
Akis Linardos \inst{14}
\and
Carla Pitarch \inst{14,15}
\and
Sanyukta Adap \inst{14}
\and
Jeffrey Rudie \inst{19} \and
Maria Correia de Verdier \inst{19, 20} \and
Rachit Saluja \inst{34, 35} \and
Evan Calabrese \inst{21} \and
Dominic LaBella \inst{22} \and
Mariam Aboian \inst{23} \and
Ahmed W. Moawad \inst{24} \and
Nazanin Maleki \inst{23} \and
Udunna Anazodo \inst{25, 36} \and
Maruf Adewole \inst{26} \and
Marius George Linguraru \inst{27, 28} \and
Anahita Fathi Kazerooni \inst{29, 30, 31} \and
Zhifan Jiang \inst{27} \and
Gian Marco Conte \inst{32} \and
Hongwei Li \inst{33} \and
Juan Eugenio Iglesias \inst{33} \and
\\
Spyridon Bakas \inst{14,15,16,17,18} 
\thanks{equal senior corresponding authors} \and
Benedikt Wiestler \inst{3,4}  \samethanks
\and \\
Marie Piraud \inst{2}
\samethanks
\and
Bjoern Menze \inst{1} \samethanks
\\
}

%

\authorrunning{F. Kofler et al.}
%

\institute{
Department of Quantitative Biomedicine, University of Zurich, Zurich, Switzerland
\and
Helmholtz AI, Helmholtz Zentrum München, Germany \and
Department of Diagnostic and Interventional Neuroradiology, School of Medicine, Klinikum rechts der Isar, Technical University of Munich, Germany \and
TranslaTUM - Central Institute for Translational Cancer Research, Technical University of Munich, Germany \and
Chair for AI in Medicine, Klinikum Rechts der Isar, Munich, Germany \and
Department of Radiation Oncology, Klinikum rechts der Isar, Technical University of Munich, Germany \and
Institute of Clinical Neuroimmunology, Faculty of Medicine, Ludwig Maximilian University, Munich, Germany \and
School of Computation, Information and Technology, Technical University of Munich, Germany  \and
Center for Digital Technologies in Dentistry and CAD/CAM, Danube Private University, Krems an der Donau, Austria \and
Department of Neurology, Clinical Neuroscience Center and Brain Tumor Center, University and University Hospital Zurich, 8091 Zurich, Switzerland
\and
School Social Sciences and Technology, Technical University of Munich, Germany 
\and
Department of Medical Radiation Physics, Stockholm University, Solna, Sweden
\and
Department of Oncology-Pathology, Karolinska Institutet, Solna, Sweden, Sweden
\and
Department of Pathology and Laboratory Medicine, Indiana University School of Medicine, Indianapolis, IN, USA
\and
Indiana University Melvin and Bren Simon Comprehensive Cancer Center, Indianapolis, IN, USA
\and
Departments of Radiology and Imaging Sciences; Neurological Surgery; Biostatistics and Health Data Science, Indiana University School of Medicine, Indianapolis, IN, USA
\and
Department of Computer Science, Luddy School of Informatics, Computing, and Engineering, Indiana University, Indianapolis, IN, USA
\and
Medical Research Group, MLCommons
\and
Department of Radiology, University of California San Diego, San Diego, CA, USA
\and
Department of Surgical Sciences, section of Neuroradiology, Uppsala University, Sweden
\and
Department of Radiology, Duke University, Durham, NC, USA
\and
Departments of Radiation Oncology, Duke University, Durham, NC, USA
\and
Department of Radiology, The Children’s Hospital of Philadelphia, PA, USA
\and
Department of Radiology, Mercy Catholic Medical Center, Darby, PA, USA
\and
Montreal Neurological Institute (MNI), McGill University, Montreal, QC, Canada
\and
Medical Artificial Intelligence Laboratory (MAI Lab), Lagos, Nigeria
\and
Sheikh Zayed Institute for Pediatric Surgical Innovation, Children’s National Hospital, Washington, D.C., USA
\and
Departments of Radiology and Pediatrics, George Washington University School of Medicine and Health Sciences, Washington, D.C., USA
\and
Center for Data-Driven Discovery in Biomedicine, Children’s Hospital of Philadelphia, Philadelphia, PA, USA
\and
Department of Neurosurgery, University of Pennsylvania, Philadelphia, PA, USA
\and
Division of Neurosurgery, Children’s Hospital of Philadelphia, Philadelphia, PA, USA
\and
Department of Radiology, Mayo Clinic, Rochester, MN, USA
\and
Athinoula A. Martinos Center for Biomedical Imaging, Massachusetts General Hospital, Boston, MA, USA
\and
Department of Electical and Computer Engineering , Cornell University and Cornell Tech, New York, NY, USA
\and
Department of Radiology, Weill Cornell Medicine, New York, NY, USA
\and
Medical Artificial Intelligence Laboratory (MAI Lab), Lagos, Nigeria
}
\maketitle              
\begin{abstract}
The Brain Tumor Segmentation (BraTS) cluster of challenges has significantly advanced brain tumor image analysis by providing large, curated datasets and addressing clinically relevant tasks. However, despite its success and popularity, algorithms and models developed through BraTS have seen limited adoption in both scientific and clinical communities.
To accelerate their dissemination, we introduce \emph{BraTS orchestrator}, an open-source Python package that provides seamless access to state-of-the-art segmentation and synthesis algorithms for diverse brain tumors from the BraTS challenge ecosystem. Available on \href{https://github.com/BrainLesion/BraTS}{GitHub}, the package features intuitive tutorials designed for users with minimal programming experience, enabling both researchers and clinicians to easily deploy winning BraTS algorithms for inference.
By abstracting the complexities of modern deep learning, \emph{BraTS orchestrator} democratizes access to the specialized knowledge developed within the BraTS community, making these advances readily available to broader neuro-radiology and neuro-oncology audiences.
\keywords{
BraTS challenge,
brain tumor,
MRI,
biomedical image analysis,
glioma, 
metastasis,
meningioma,
pediatric tumor,
segmentation,
inpainting,
synthesis
}

\end{abstract}

\section{Introduction}
\label{sec:introduction}

Brain tumors, though infrequent, cause substantial mortality and morbidity across all age demographics \citep{BrainTumorEpidemiology}.
These tumors encompass a diverse group of neoplasms, which can be roughly subdivided into \textit{primary} brain tumors, i.e., those tumors that originate in the brain, and \textit{secondary}, metastatic tumors that spread into the brain. Consequently, the genetic, histological, and ultimately also MR imaging landscape of these tumors is starkly heterogeneous: For instance, \eac{MEN}, typically benign tumors arising from arachnoidal cap cells, are characterized by slow growth patterns and often solid contrast enhancement.
Conversely, \textit{glioblastoma}, the most prevalent and aggressive malignant primary brain tumors in adults, carry a poor prognosis, often with a survival of less than 14 months \citep{baid2021rsnaasnrmiccaibrats2021benchmark} and have an inhomogeneous MRI appearance, often containing both large necrotic and vital, strongly contrast-enhancing areas as well as large peritumoral edema.
Current treatment protocols for brain tumors are contingent upon tumor type but generally involve a combination of surgical resection, \eac{RT}, and chemotherapy.

\eac{MRI}, specifically \eac{mpMRI}, serves as the standard imaging modality for brain tumor assessment, providing high-resolution visualization of tumor tissues.
These images are critical for diagnostic purposes, \eac{RT} planning, and disease monitoring.
The inherent heterogeneity of brain tumor genotypes, reflected in the radiological characteristics observed through \eac{MRI}, manifests itself as tumor subregions that exhibit heterogeneous intensity, texture, and shape.
The variability in these tumors makes their manual delineation a complex and time-consuming task.

Numerous computational models for brain tumor segmentation have been developed using private datasets.
However, benchmarking and comparative performance analysis of these strategies have been hampered by several factors, including variations in imaging modality (e.g., structural \eac{MRI},  diffusion,
and functional \eac{MRI}), brain tumor type (e.g., pathological diagnosis, tumor grade, and growth pattern), and disease state (e.g., pre-surgery, post-surgery pre-\eac{RT}, post-\eac{RT}).

To address these limitations, the \eac{BraTS} was established in 2012 at the \eac{MICCAI} to facilitate the objective evaluation of state-of-the-art algorithms for automated brain tumor segmentation in \eac{mpMRI} \citep{menze2014multimodal}.
\eac{BraTS} has subsequently become a driving force in advancing image-based brain tumor analysis by providing a standardized benchmarking environment and curated datasets.
Specifically, for each \eac{MICCAI} conference since its inception, \eac{BraTS} organizers have provided a curated set of structural \eac{mpMRI} scans, including \eac{T1n}, \eac{T1c}, \eac{T2w}, and \eac{FLA} images, along with corresponding expert-generated segmentation masks \citep{CancerGenoeAtlasBakas2017}.
While the initial focus of the \eac{BraTS} challenge for its first nine years was the segmentation of pre-operative \eac{GLI} into the tumor subregions (\eac{ET}, \eac{NETC}, and \eac{ED}) \citep{bakas2018identifying}, its scope has broadened in recent years.
This expansion now encompasses a wider range of segmentation challenges, including post-treatment \eac{GLI} \citep{deverdier20242024braintumorsegmentationadultgliomaposttreatment}, \eac{MEN} (both pre-and post-\eac{RT}) \citep{labella2024braintumorsegmentationbratsMeningioma} \citep{labella2023asnrmiccaibraintumorsegmentationMeningioma}, \eac{PED} tumors \citep{kazerooni2024braintumorsegmentationpediatrics24}, \eac{METS} \citep{moawad2024braintumorsegmentationbratsmets}, \eac{SSA} adult \eac{GLI} \citep{adewole2023braintumorsegmentationbratsafrica}, \eac{GoAT} \href{https://www.synapse.org/Synapse:syn53708249/wiki/627506} and tasks beyond segmentation, such as missing MRI synthesis \citep{li2024braintumorsegmentationbratsbrasyn}, local synthesis via tissue inpainting \citep{kofler2024braintumorsegmentationbratsinpainting}, and assessment of the heterogeneous histologic landscape of \eac{GLI} \citep{bakas2024bratspathchallengeassessingheterogeneous}. BraTS now addresses pertinent clinical challenges across tumor entities and disease courses, making BraTS the central driving force for brain tumor image analysis model development and benchmarking.


The advancement of \eac{DL} techniques has facilitated the development of robust computational algorithms for automated brain tumor segmentation.
Inspired by the groundbreaking U-Net \citep{ronnenbergerUnet2015} and V-Net architectures \citep{MilletariVNet}, a multitude of encoder-decoder models have been developed for diverse segmentation tasks.
These models primarily introduce innovations through modifications to network architecture and/or optimization procedures.
Several \eac{CNN} and transformer-based models such as U-Net++ \citep{NestedUnet}, SegResNet \citep{MyronenkoSegResNet}, nn-UNet \citep{isensee2021nnu} \citep{isensee2024nnunetrevisitedrigorousvalidation}, nnFormer \citep{zhou2022nnformerinterleavedtransformervolumetric}, and Swin UnetR \citep{HatamizadehSwinUnetR} have recently demonstrated superior performance in brain tumor segmentation.

The developed robust segmentation models have provided a foundation for other diagnostic and prognostic applications in brain tumors.
Retrospective analysis of a large, multi-institutional dataset of \eac{GLI} patients has shown that objective assessment of tumor response to treatment, using a baseline segmentation model, surpasses human evaluation in predicting patient survival \citep{KoflerSurvivalGLI} \citep{Kickingereder2019-kz}, indicating the potential value of integrating these algorithms into clinical practice.
Accordingly, \eac{BraTS} has facilitated a range of image-based brain tumor analyses extending beyond segmentation, encompassing synthetic tumor generation \citep{LiuSyntheticMaskBraTS23} brain tumor growth modeling \citep{BakasGliomaGrowth2020} \citep{SarthakGliomaGrowthDL2021} \citep{Learn-Morph-Infer2023}, molecular subtype classification \citep{Akbari2024-gm} \citep{SubtypePredictionReview}, treatment response assessment \citep{RanoAI1-2024} \citep{RanoAI2-2024}, and prognostic and survival predictions \citep{Fathi_Kazerooni2022-lb} \citep{Fathi_Kazerooni2025-ci} \citep{LiuPrognostication2024}.

Despite the robust and accurate solutions demonstrated in the \eac{BraTS} challenges, the clinical translation of these algorithms remains limited.
While issues such as data privacy and ethical considerations are beyond the scope of this paper, technical obstacles, including data curation, conversion, preprocessing, and the development of harmonized segmentation pipelines, impede the successful deployment of these algorithms.
Current solutions are often fragmented into independent steps, hindering their unification into a cohesive and standardized workflow \citep{brain-CaPTk2018} \citep{pati2020cancer} \citep{kofler2020brats}.
A centralized approach offers significant potential to facilitate broader implementation of advancements in image analysis.
Therefore, this paper introduces \href{https://github.com/BrainLesion/BraTS}{\textit{BraTs orchestrator}}, a centralized and standardized framework designed to improve the accessibility and executability of advanced brain tumor analysis \eac{DL} models.

\section{BraTS orchestrator}
\label{sec:orchestrator}
\textit{BraTs orchestrator} represents an effort focused on the creation of an accessible, Python \eac{API}-driven pipeline and adheres to the principles of open and reproducible science; thus, it is distributed under the \href{https://www.apache.org/licenses/LICENSE-2.0}{Apache License, Version 2.0}.
This package also efforts to support proprietary operating systems, such as Microsoft Windows and Mac OSX, beyond Linux.
This broader compatibility is pursued wherever technically feasible, acknowledging the varied and sometimes restrictive computing environments prevalent in clinical and research institutions.
This is facilitated through \href{https://www.docker.com/}{Docker} containerization
technology.
It continues and simplifies the previously established \textit{BraTS Toolkit} segmentation module \cite{kofler2020brats} that similarly provided access to top algorithms from older \eac{BraTS} challenges.

The primary goal of this effort is to improve the availability and operational ease of algorithms that have demonstrated winning performance in the \eac{BraTS} challenge.
More precisely, this effort provides a minimalist and intuitive Python \eac{API} tool that enables users to run the winning algorithms from recent iterations of the \eac{BraTS} challenges, specifically during their inference phase.

In addition, the \textit{BraTS Fusionator} module, developed within the previously established \href{https://github.com/neuronflow/BraTS-Toolkit}{BraTS Toolkit} \cite{kofler2020brats}, can be employed to combine and ensemble the candidate segmentations generated by these algorithms.
This is achieved through the application of fusion methods such as majority voting and iterative SIMPLE fusion, ultimately yielding consensus segmentations.

Furthermore, by leveraging functionalities derived from the
\href{https://github.com/BrainLesion/preprocessing}{BrainLesion suite preprocessing package}, \textit{BraTs orchestrator} enables a modular workflow.
This workflow is dedicated to the preprocessing of both single-channel \eac{MRI} and \eac{mpMRI} datasets, thereby ensuring their compatibility with the input specifications required by the trained models.



\subsection{\eac{BraTS} challenges}
\label{sec:brats_challenge}
The established convention for participation in the \eac{BraTS} challenge grants participants access to training data, inclusive of the corresponding reference labels, commonly referred to as ground truth. 
In addition, participants receive a validation dataset, which, however, does not include these reference labels.
This framework allows participants to develop and optimize their computational models utilizing both the training and validation datasets. 
For the final testing phase, participants are required to containerize their developed solutions according to specified protocols.
These containerized solutions are then evaluated by the challenge organizers on unseen testing data.
The ranking of submitted models is typically determined through a comparative analysis of quantitative performance metrics, such as lesion-wise \eac{DSC}, \eac{HD}, and recently \eac{NSD} metrics for segmentation tasks, applied to tumoral subregions and their combinations.
This quantitative assessment is subsequently complemented by extensive statistical analysis.

The scope of the \textit{BraTs orchestrator} is limited to challenges centered on the \eac{MRI} data, thus excluding the task related to histology due to the fundamental differences inherent in histological imaging modalities. Currently, \textit{BraTs orchestrator} incorporates seven segmentation tasks and two synthesis tasks about various brain tumor entities.

The primary objective of the segmentation tasks is to automatically segment different types of brain tumors, including \eac{GLI}, \eac{METS}, \eac{MEN}, \eac{PED}, and \eac{SSA} into their constituent subregions. These include, for example, \eac{ET}, \eac{NETC}, \eac{SNFH}, \eac{CC}, and \eac{RC}, as well as the overall tumor region, commonly referred to as the \eac{GTV}.

The \eac{MRI} data utilized in the \eac{BraTS} challenges underwent preprocessing steps including registration to atlas space.
Within these challenges, two distinct atlases were employed for this registration: MNI152 \citep{MniAtlas2009Paper} and SRI24 \citep{rohlfing2010sri24}.
It is important to note that although multiple versions of these atlases are available, the precise MNI152 and SRI24 versions utilized in the \eac{BraTS} challenges will be made available through the preprocessing modules of \textit{BraTS Orchestrator}.
\Cref{tab:OrchestratorSegTasks} provides a comprehensive summary of the segmentation tasks within the \eac{BraTS} cluster of challenges.

\begin{table}[htbp]
  \caption{
  \textit{BraTS orchestrator} features algorithms from the BraTS segmentation challenges summarized in this table.
  Moreover, missing modality synthesis and brain lesion inpainting algorithms are included, c.f. \Cref{fig:brats_tasks}.
  }
  \label{tab:OrchestratorSegTasks}
  \scriptsize
  \begin{tabularx}{\textwidth}{@{}ccCCcc@{}}
    \toprule
    \textbf{Task}                                                     &
    \textbf{Year(s)}                                                  &
    \textbf{Inputs}                                                   &
    \textbf{Preprocessing}                                            &
    \textbf{Spatial Space}                                            &
    \textbf{Tumor Labels}                                               \\
    \midrule
    \eac{GLI}-pre                                                     &
    2023                                                              &
    \eac{mpMRI} \par(\eac{T1c}, \eac{T1n},\par \eac{T2w}, \eac{FLA})  &
    co-registration \par skull stripping \par atlas registration      &
    SRI24                                                             &
    \eac{ET}, \eac{NETC}, \eac{SNFH}                                    \\
    \midrule
    \eac{GLI}-post                                                    &
    2024                                                              &
    \eac{mpMRI} \par(\eac{T1c}, \eac{T1n},\par \eac{T2w}, \eac{FLA})  &
    co-registration \par skull stripping \par atlas registration       &
    MNI152                                                            &
    \eac{ET}, \eac{NETC}, \eac{SNFH}, \eac{RC}                          \\
    \midrule
    \eac{SSA}                                                         &
    2023 -- 2024                                                      &
    \eac{mpMRI} \par (\eac{T1c}, \eac{T1n},\par \eac{T2w}, \eac{FLA}) &
    co-registration\par skull stripping \par atlas registration         &
    SRI24                                                             &
    \eac{ET}, \eac{NETC}, \eac{SNFH}                                    \\
    \midrule
    \eac{MEN}-pre                                                     &
    2023                                                              &
    \eac{mpMRI} \par (\eac{T1c}, \eac{T1n},\par \eac{T2w}, \eac{FLA}) &
    co-registration\par skull stripping\par atlas registration        &
    SRI24                                                             &
    \eac{ET}, \eac{NETC}, \eac{SNFH}                                    \\
    \midrule
    \eac{METS}                                                        &
    2023                                                              &
    \eac{mpMRI} \par (\eac{T1c}, \eac{T1n},\par \eac{T2w}, \eac{FLA}) &
    co-registration\par skull stripping\par atlas registration         &
    SRI24                                                             &
    \eac{ET}, \eac{NETC}, \eac{SNFH}                                    \\
    \midrule
    \eac{PED}                                                         &
    2023 -- 2024                                                      &
    \eac{mpMRI} \par (\eac{T1c}, \eac{T1n},\par \eac{T2w}, \eac{FLA}) &
    co-registration\par defacing                                      &
    native                                                            &
    \eac{ET}, \eac{NETC}, \eac{CC}, \eac{ED}                            \\
    \midrule
    \eac{GoAT}                                                        &
    2024                                                              &
    \eac{mpMRI} \par (\eac{T1c}, \eac{T1n},\par \eac{T2w}, \eac{FLA}) &
    co-registration\par skull stripping \par atlas registration        &
    SRI24                                                             &
    \eac{ET}, \eac{NETC}, \eac{SNFH}                                    \\
    \midrule
    \eac{MEN}-rt                                                      &
    2024                                                              &
    \eac{T1c}                                                         &
    defacing                                                          &
    native                                                            &
    \eac{GTV}                                                           \\
    \bottomrule
  \end{tabularx}
\end{table}


Furthermore, beside segmentation, \textit{BraTS orchestrator} includes algorithms for the following two image synthesis tasks:

\begin{enumerate}

\item \textbf{Local Synthesis of Tissue via Inpainting}: This non-segmentation task focuses on the generation of synthetic, healthy-appearing brain tissue in \eac{MRI} scans that originally depict brain tumors, utilizing inpainting methodologies.
The dataset utilized for this task was derived from the \eac{GLI}-pre task, encompassing data for training, validation, and testing. 
Additionally, a second test set from the \eac{MEN}-pre task was incorporated to evaluate the generalization capabilities of the algorithms.

\item \textbf{Missing \eac{MRI} Synthesis}: Analogous to the preceding task, this non-segmentation challenge is concerned with the synthesis of anatomically plausible \eac{MRI} sequences that are missing from a given dataset.
The performance of numerous segmentation models is critically dependent on the availability of all four standard \eac{mpMRI} sequences.
However, in routine clinical practice, it is not uncommon for only a subset of these sequences to be acquired.

\end{enumerate}

\begin{figure}
    \centering
    \includegraphics[width=\linewidth]{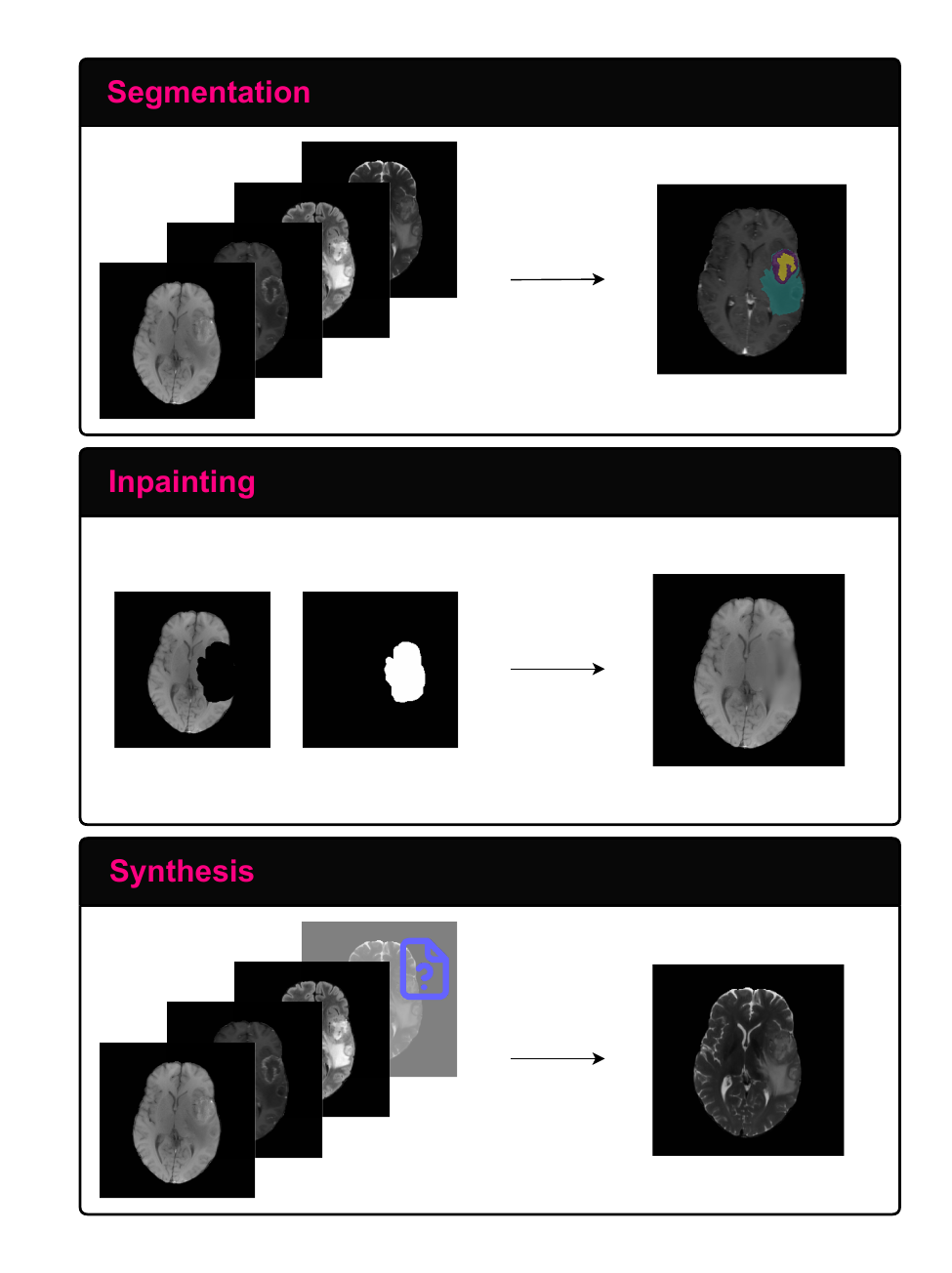}
    \caption{
    \textit{BraTS orchestrator} features algorithms for tumor segmentation, missing  \eac{MRI} modality synthesis and brain lesion inpainting.
    The figure illustrates the respective inputs to the algorithms on the left and the computed outputs on the right.
    \Cref{fig:brats_segmentations} illustrates the particularities of the individual BraTS segmentation tasks.
    }
    \label{fig:brats_tasks}
\end{figure}

\begin{figure}
    \centering\includegraphics[width=\linewidth]{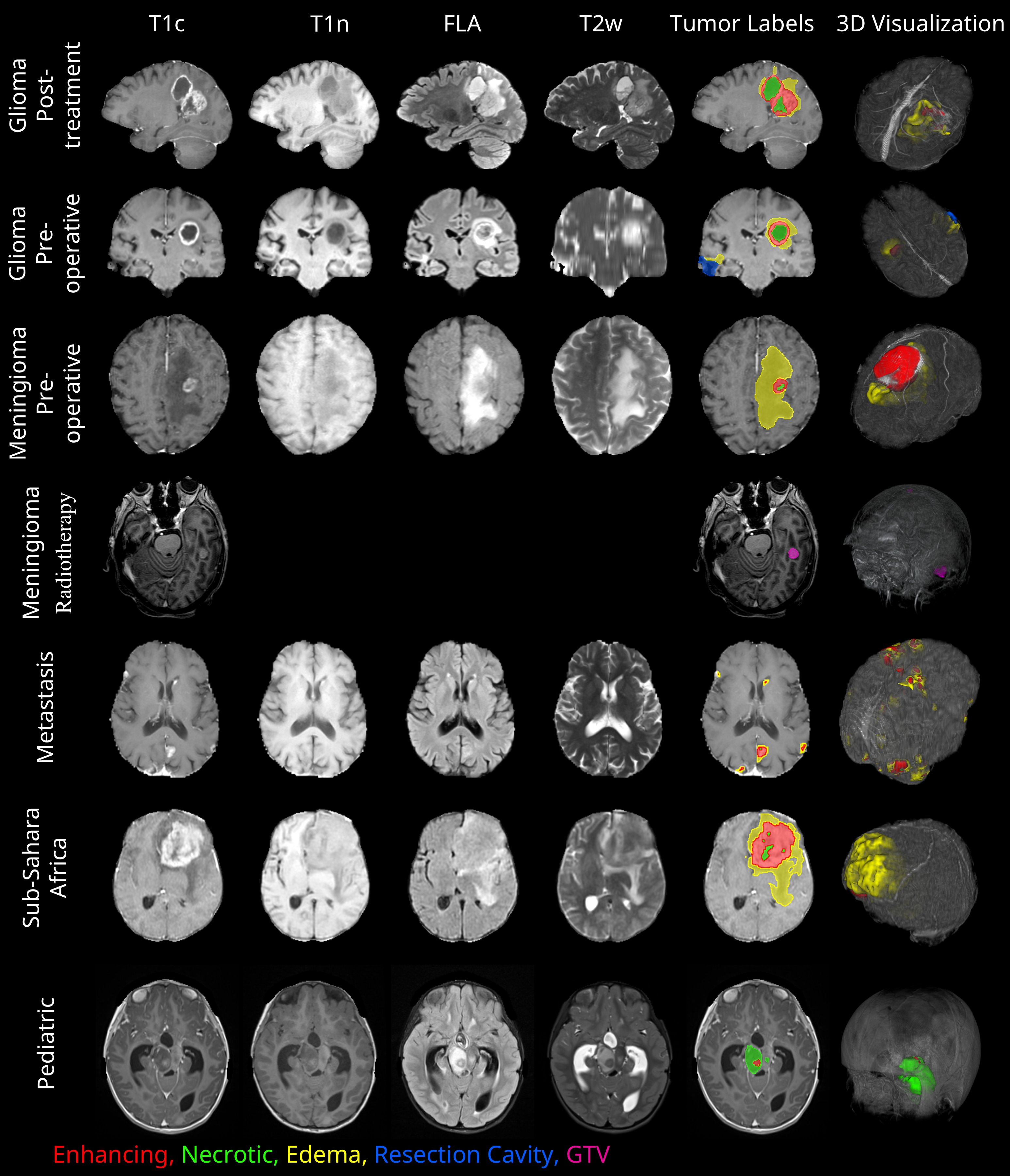}
    \caption{
    Illustration of the different segmentation tasks for diverse brain tumor types within the \eac{BraTS} segmentation challenges.
    Each row represents a distinct segmentation task, with the first four columns indicating the required \eac{MRI} data.
    The segmentation labels for various tumor subregions, including \eacf{ET}, \eacf{NETC}, \eacf{ED}, \eacf{RC}, and \eacf{GTV}, are color-coded and overlaid on both 2D slices and 3D visualizations.
    }
    \label{fig:brats_segmentations}
\end{figure}

\autoref{fig:brats_tasks} provides a general visual representation of the described tasks, indicating their respective inputs and outputs. Subsequently, \autoref{fig:brats_segmentations} details the segmentation tasks within the \eac{BraTS} challenge by illustrating the requisite \eac{MRI} sequences and corresponding segmentation labels for each task.

\subsection{Algorithms from \eac{BraTS}}
\label{sec:brats_algorithms}
Recent outcomes from the \eac{BraTS} challenges demonstrate that leading algorithms frequently employ established and robust segmentation models, including but not limited to nnU-Net \citep{isensee2024nnunetrevisitedrigorousvalidation}, Swin UNETR \citep{HatamizadehSwinUnetR}, MedNeXt \citep{dkfz-mednext}, SegResNet \citep{MyronenkoSegResNet}, Auto3DSeg \citep{myronenko2023automated3dsegmentationkidneys}, and STU-Net \citep{huang2023stunetscalabletransferablemedical}. 
Characteristically, these robust models are utilized in conjunction with innovative modifications to their architectures, objective functions, and data augmentation techniques, as well as customized preprocessing and postprocessing strategies. 
\Cref{tab:brats23summary,tab:brats24summary} summarize the methodologies implemented by the winning algorithms of the \eac{BraTS} 2023 and \eac{BraTS} 2024 challenges across diverse tasks.

\afterpage{%
    \clearpage
    \centering
    \begin{landscape}
        \begin{table}[htbp]
            \caption{Summary of the top-performing algorithms in BraTS 2023 segmentation challenges. IN refers to intensity normalization.}
            \label{tab:brats23summary}
            \tiny
            \begin{tabularx}{\linewidth}{L{1.2cm}XCCCCM{1.0cm}M{0.8cm}CM{0.8cm}C}
                \toprule
                \textbf{Task}                         &
                \textbf{Reference}                    &
                \textbf{Pre-processing}               &
                \textbf{Architecture}                 &
                \textbf{Loss}                         &
                \textbf{Patch size}                   &
                \textbf{Batch size}                   &
                \textbf{Pre-trained}                  &
                \textbf{Aug-mentation}                &
                \textbf{TTA}                          &
                \textbf{Post-processing}             \\ 
                \midrule
                \multirow{3}{*}{GLI-pre}              &
                \citet{ferreira2024wonbrats2023adult} & default nnU-Net                        & nnU-Net, Swin UNETR & batch dice                           & 128x160x112, 128x128x128              & 5                     & \xmark & registration, GANs, rotation, scaling, brightness, gamma, elastic deformation                           & \cmark                                                                              & model ensembling, thresholding                                                                                       \\ \gcmidrule{11}
                                                      & \citet{MyronenkoBraTS23}                           & N/A                 & N/A                                  & N/A                                   & N/A                   & N/A    & N/A                                                                                                     & N/A                                                                                 & N/A                                                 & N/A                                                            \\ \gcmidrule{11}
                                                      & \citet{Maani-GLI23}                                & Cropping, IN        & MedNeXt, SegResNet                   & batch dice + focal loss               & 128x128x128           & 2      & \xmark                                                                                                  & flipping, shifting, intensity scaling                                               & \cmark                                              & fold ensembling, thresholding, connected component, clustering \\
                \midrule
                \multirow{3}{*}{MEN-pre}              
                & \citet{MyronenkoBraTS23}                           & N/A                 & N/A                                  & N/A                                   & N/A                   & N/A    & N/A                                                                                                     & N/A                                                                                 & N/A                                                 & N/A                                                            \\ \gcmidrule{11}
                &\citet{HuangMenPre23}                              & default nnU-Net                        & STU-Net             & default nnU-Net                      & 128x128x128                           & 2                     & \cmark & brightness, gamma, rotation, scaling, mirror, elastic deformation                                       & \cmark                                                                              & N/A                                                                                       \\ \gcmidrule{11}
                                                      
                                                      & \citet{CapellanMartin2024ModelEF}                          & default nnU-Net     & nnU-Net, Swin UNETR                  & default nnU-Net; dice + focal loss    & 128x160x112, 96x96x96 & N/A    & \xmark                                                                                                  & default nnU-Net                                                                     & \cmark                                              & model ensembling, connected component, thresholding            \\
                \midrule
                \multirow{3}{*}{METs}     
                & \citet{MyronenkoBraTS23}                          & N/A                 & N/A                                  & N/A                                   & N/A                   & N/A    & N/A                                                                                                     & N/A                                                                                 & N/A                                                 & N/A                                                            \\ \gcmidrule{11}
                &\citet{YangMets23}                               & cropping, IN                           & 3D-TransUNet        & cross entroy + dice + classification & 128x128x128                           & 2                     & \cmark & rotation, scaling, flipping, blurring Gaussian noise, color jittering, gamma, low-resolution simulation & \cmark                                                                              & fold ensembling                                                                                                      \\ \gcmidrule{11}
                                                      
                                                      & \citet{HuangMenPre23}                               & default nnU-Net     & STU-Net                              & default nnU-Net                       & 128x128x128           & 2      & \xmark                                                                                                  & brightness, gamma, rotation, scaling, elastic deformation, mirror                   & \cmark                                              & N/A                                 \\
                \midrule
                \multirow{3}{*}{SSA}        
                & \citet{MyronenkoBraTS23}                           & N/A                 & N/A                                  & N/A                                   & N/A                   & N/A    & N/A                                                                                                     & N/A                                                                                 & N/A                                                 & N/A                                                            \\ \gcmidrule{11}
                &\citet{AmodSSA23}                         & cropping, IT, added foreground channel & Optimized U-Net     & binary cross-entropy + dice loss     & 128x128x128                           & N/A                   & \cmark & flipping, brightness, Gaussian noise, blurring, contrast adjustments                                    & \cmark                                                                              & fold ensembling, connected component, thresholding                                                                   \\ \gcmidrule{11}
                                                      
                                                      & \citet{HuangMenPre23}                              & default nnU-Net     & STU-Net                              & default nnU-Net                       & 128x128x128           & 2      & \xmark                                                                                                  & brightness, gamma, rotation, scaling, elastic deformation, mirror                   & \cmark                                              & N/A                                 \\
                \midrule
                \multirow{3}{*}{PEDs}                 &
                \citet{CapellanMartin2024ModelEF}           & default nnU-Net                        & nnU-Net, Swin UNETR & default nnU-Net; dice + focal loss   & 128x160x112, 96x96x96                 & N/A                   & \xmark & default nnU-Net                                                                                         & \cmark                                                                              & model ensembling, connected component, thresholding                                                                  \\ \gcmidrule{11}
                                                      & \citet{MyronenkoBraTS23}                           & N/A                 & N/A                                  & N/A                                   & N/A                   & N/A    & N/A                                                                                                     & N/A                                                                                 & N/A                                                 & N/A                                                            \\ \gcmidrule{11}
                                                      & \citet{ZhouPEDS23}                              & default nnU-Net     & nnU-Net                              & cross entropy + adaptive regions loss & 128x128x128           & 2      & \cmark                                                                                                  & rotation, scaling, Gaussian noise, gamma, blurring, brightness contrast adjustments & \cmark                                              & fold ensembling                                                \\
                \bottomrule
            \end{tabularx}
        \end{table}
    \end{landscape}
}

\afterpage{%
    \clearpage
    \centering
    \begin{landscape}
    \begin{table}[htbp]
        \caption{Summary of the top-performing segmentation algorithms in BraTS 2024 challenges. IN refers to intensity normalization.}
        \label{tab:brats24summary}
        \tiny
        \begin{tabularx}{\linewidth}{L{1.2cm}XCCCCM{1.0cm}M{0.8cm}CM{0.8cm}C}
            \toprule
            \textbf{Task} &
              \textbf{Reference} &
              \textbf{Pre-processing} &
              \textbf{Architecture} &
              \textbf{Loss} &
              \textbf{Patch size} &
              \textbf{Batch size} &
              \textbf{Pre-trained} &
              \textbf{Aug-mentation} &
              \textbf{TTA} &
              \textbf{Post-processing} \\ \midrule
              \multirow{3}{*}{GLI-post} 
              &\citet{ferreira2024improvedmultitaskbraintumour}&default nnU-Net	&nnU-Net, MedNeXt,  Swin UNETR	&default nnU-Net	&default nnU-Net 	&default nnU-Net	&\xmark&	registration, GANs, default nnU-Net&	\cmark&	region-based thresholding, model ensembling, \\ \gcmidrule{11}
              &  \citet{kim2024effectivesegmentationposttreatmentgliomas} &additional input channel, default nnU-Net	&nnU-Net, SegResNet	&default nnU-Net	&128x160x112, 160x192x160	&2, 3	& \xmark&	default nnU-Net	&\cmark&	model ensembling \\ \gcmidrule{11}
              &\citet{celaya2024mistsimplescalableendtoend}&Cropping, IN, resampling, reorientation	&nnU-Net	&Dice + CE	&128x128x128	&32	&\xmark&	default  Nvidia data loading library (DALI)&	\cmark&	Fold ensembling \\ 
              \midrule
              \multirow{3}{*}{MEN-rt} 
              &\citet{AbramovaBraTS24}	&target and boundary segmentation, default nnU-Net	&nnU-Net	&default nnU-Net	&96x160x160	&default nnU-Net	&\xmark&	default nnU-Net	&\cmark&	Thresholding, Model ensemble  \\ \gcmidrule{11}
              &\citet{AstarakiBraTS24}	&Cropping, intensity clipping, default nnU-Net	&nnU-Net, MedNeXt,  SegResNet	&default nnU-Net	&112x160x128 128x128x128	&4, 2	&\xmark&	default nnU-Net	&\cmark&	fold ensembling \\ \gcmidrule{11}
              &\citet{ferreira2024improvedmultitaskbraintumour}&default nnU-Net	&nnU-Net, MedNeXt	&default nnU-Net	&default nnU-Net	&default nnU-Net	&\xmark&	registration, GANs, default nnU-Net	&\cmark&	region-based thresholding, model ensembling \\
              \midrule
              \multirow{3}{*}{SSA} 
              &\citet{parida2024adultgliomasegmentationsubsaharan}&Unsupervised fold split,  default nnU-Net 	&nnU-Net, MedNeXt	&default nnU-Net	&128x128x128	&default nnU-Net	&\cmark&	default nnU-Net	&\cmark&	region-based thresholding, model ensembling  \\ \gcmidrule{11}
              &\citet{zhao2024transferringknowledgehighqualitylowquality} &Super-resolution,  default nnU-Net	&Modified nnU-Net	&Batch Dice + CE&	128x128x128	&default nnU-Net	&\cmark	&brightness, gamma, rotation, scaling, elastic deformation, gamma	&\cmark&	model ensembling \\ \gcmidrule{11}
              &\citet{hashmi2024optimizingbraintumorsegmentation} &default nnU-Net	&MedNeXt	&Dice + Focal	&128x160x112	&2	&\cmark&	flipping, intensity scaling, shifting	&\cmark&	Thresholding, model ensembling \\
              \midrule
              \multirow{3}{*}{PEDs} 
              &\citet{AstarakiBraTS24}	&Cropping, intensity clipping, default nnU-Net	&nnU-Net, MedNeXt,  SegResNet	&default nnU-Net	&112x160x128 128x128x128	&4, 2	&\xmark&	default nnU-Net	&\cmark&	fold ensembling \\ \gcmidrule{11}
              &\citet{MulvanyBraTS24}	&default nnU-Net	&nnU-Net - cascade	&default nnU-Net	&default nnU-Net	&2	&\xmark&	default nnU-Net	&\cmark&	model ensembling \\ \gcmidrule{11}
              &\citet{hashmi2024optimizingbraintumorsegmentation}&default nnU-Net	&MedNeXt	&Dice + Focal	&128x160x112	&2&	\xmark&	flipping, intensity scaling, shifting	&\cmark&	Thresholding, model ensembling \\
              \midrule
              \multirow{1}{*}{GoAT} 
              &\citet{NiuBraTS24}	&default nnU-Net	&nnU-Net, mean teacher and alignment	&Unified focal 	&160x192x160	&4	&\xmark&	rotation, noise addition, intensity shift	&?	&class-wise thresholds, ensembling \\
              \bottomrule
        \end{tabularx}
    \end{table}
    \end{landscape}
}

\section{Applications and tutorials}
\label{sec:applications}
The modular design of the \textit{BraTS orchestrator} facilitates the development of customizable pipelines, adaptable to a range of applications such as integration of post-processing techniques and ensembling methodologies.
The provided tutorials on the \textit{BraTS orchestrator}
\href{https://github.com/BrainLesion/BraTS}{\textit{GitHub repository}}
Illustrate the application of its segmentation and synthesis modules during the inference phase.
Nevertheless, the inherent flexibility of this framework supports a broader spectrum of use cases.
Therefore, this section presents a selection of common applications within the neuro-radiology and neuro-oncology fields for which the \textit{BraTS orchestrator} offers a versatile and extensible modular platform.

\begin{itemize}
    \item \textbf{Native Space Segmentation} Although most \eac{BraTS} segmentation tasks require preprocessed image data, including atlas registration— a technique established to reduce inter-dataset variations and improve model efficacy— clinical applications, such as \eac{RT} planning, necessitate tumor segmentation directly within the native image space.
          This requirement implies that the segmentation masks generated by the \textit{BraTS orchestrator} must undergo a precise inverse transformation to their original spatial orientation.
          This transformation is feasible by storing the forward atlas registration transformation matrices.
          Subsequently, these matrices are utilized to apply an inverse warp to the resultant segmentation masks.
          This strategy effectively combines the benefits of robust segmentation achieved by \eac{BraTS}-winning algorithms on preprocessed data with the imperative of clinical compatibility in the native image space.
          Future updates of the \textit{BraTS orchestrator} are planned to support native space segmentations.
    \item \textbf{DICOM Support} While \eac{DICOM} is the standard for storing and transmitting medical images, it often comes with a complex and extensive metadata structure that can vary significantly between manufacturers and acquisition protocols.
          This variability makes programmatic access and consistent interpretation challenging for research applications focused solely on image processing and analysis.
          \eac{NIfTI}, on the other hand, provides a simpler, more standardized, and research-friendly file format.
          It consolidates the essential image data and spatial transformation information into a single, compact header, making it much easier for research and development purposes.
          Furthermore, \eac{NIfTI} is widely supported by a vast ecosystem of open-source neuroimaging and medical image analysis tools and libraries (e.g., ANTs, FSL).

          The \textit{BraTS orchestrator} mandates \eac{NIfTI} file formats for input image data and subsequently yields results in the same format.
          However, clinical applications, particularly integration with \eac{PACS}, necessitate storing final results in the standardized \eac{DICOM} file format.
          Future iterations of the \textit{BraTS orchestrator} are planned to incorporate native \eac{DICOM} support to bridge this gap.

\end{itemize}

\section{Discussion}
\label{sec:discussion}
We introduce \textit{BraTS orchestrator}, a tool to facilitate the creation and deployment of brain tumor image analysis pipelines.
\textit{BraTS orchestrator} lowers the access barriers for both clinicians and researchers.
It therefore represents an important milestone in the dissemination and democratization of state-of-the-art image analysis algorithms.

The advancement of \eac{DL} techniques in recent years has significantly propelled the development of robust solutions for important clinically relevant tasks within medical image analysis, particularly evident in areas such as brain tumor analysis from \eac{MRI} data.
However, the translation of these methodological advancements into widespread clinical and scientific practice continues to face considerable challenges.
These impediments are multifaceted, encompassing, but not limited to, inconsistencies in target definitions for segmentation tasks, variability in data preparation and preprocessing methodologies, and a lack of standardized performance evaluation metrics.
To mitigate these inconsistencies and facilitate comparative performance assessment in biomedical image analysis, international benchmarking competitions, such as the \eac{BraTS} challenge, have emerged as indispensable platforms.

Analysis of the results achieved by the winning \eac{BraTS} models in recent years indicates that these algorithms can achieve segmentation accuracy comparable to that of expert neuroradiologists.
For instance, the winning model of \eac{BraTS} 2019, which employed a two-stage cascaded U-Net for coarse-to-fine segmentation of tumor subregions, reported average subject-wise Dice scores of 0.832, 0.836, and 0.887 for enhancing tumor \eac{ET}, \eac{NETC}, and \eac{WT}, respectively \citep{JiangBraTSwinner2019}.
These performance metrics were substantially improved in \eac{BraTS} 2023, where the winning team leveraged a large synthetic dataset for training and implemented an ensemble strategy combining three segmentation models, yielding Dice scores of 0.846, 0.876, and 0.929 for the corresponding regions \citep{ferreira2024wonbrats2023adult}.
While these results are primarily presented to highlight the increasing robustness of the developed solutions, a direct comparison between the \eac{BraTS} 2019 and 2023 outcomes necessitates careful consideration due to two principal factors: 1) the training dataset size for \eac{BraTS} 2023 was approximately three times larger than that of 2019; and 2) crucially, the \eac{BraTS} 2019 dataset was annotated de novo by experienced radiologists and subsequently corrected by expert neuro-radiologists.
Conversely, the annotation of the \eac{BraTS} 2023 dataset originated from initial segmentations generated by robust automatic \eac{DL} models, which were then corrected by expert radiologists.
This reliance on algorithmically generated initial segmentations may introduce a systemic bias in the correction process, potentially leading to an artificially higher degree of similarity between the reference standard and the model predictions \citep{kofler2023panopticainstancewiseevaluation, kofler2023approachingpeakgroundtruth}.

Despite the methodological advancements within the open-source research communities, a significant disparity persists between cutting-edge research and its widespread adoption in both scientific and clinical settings.
This disjunction arises from several interconnected factors.
For example, the intricate installation and execution requirements of many advanced \eac{DL} solutions, including the \eac{BraTS} winning algorithms, frequently necessitate specialized computational infrastructure, a high level of technical expertise in programming, and meticulous configuration of software environments.
Such prerequisites represent a substantial barrier for researchers and clinicians who may not possess the requisite programming skills.

\textit{BraTS orchestrator} is developed as an open-source, user-friendly Python package to overcome these challenges.
It provides streamlined access to state-of-the-art segmentation and synthesis winning algorithms sourced from the \eac{BraTS} cluster of challenges and enables users with minimal programming knowledge from both scientific and clinical domains to orchestrate inference with leading \eac{BraTS} algorithms.
In essence, the \eac{BraTS} Orchestrator endeavors to democratize and disseminate the knowledge distilled from the specialized \eac{BraTS} community to a broader neuro-radiology and neuro-oncology audience.

\noindent\textbf{Limitations:}
\textit{BraTS orchestrator} relies on Docker containerization, which facilitates consistent deployment and scalability of algorithms during inference, but also introduces a limitation.
This dependency necessitates the installation of Docker, which requires administrative permissions and may be restricted in certain computational environments due to security protocols.

\noindent\textbf{Outlook:}
The ultimate goal of the \textit{BraTS orchestrator} is the effective dissemination of advanced brain tumor image analysis solutions to both scientific and clinical applications.
The prospective development roadmap for the \textit{BraTS orchestrator} therefore prioritizes the development of a seamless, end-to-end image analysis workflow.
This future workflow is envisioned to be easily adaptable for integration into existing scientific and clinical pipelines, critically including capabilities for native space image analysis and \eac{DICOM} standard support.
Additionally, while the current workflow necessitates a minimum level of programming proficiency, future iterations might introduce a \eac{GUI}.
A \eac{GUI} would eliminate the need for coding, thereby significantly enhancing user accessibility and promoting the broader dissemination of these advanced models.

\vspace{\baselineskip}


\section{Acknowledgment}
\label{sec:acknowledgment}
This preprint represents a preliminary version and is subject to further development and updates.
The authorship list is currently incomplete. 
We are in the process of contacting all winning teams from the BraTS 2023 and BraTS 2024 segmentation challenges with invitations to contribute as co-authors.
We express our gratitude to all algorithm authors and challenge organizers without whom BraTS Orchestrator would not be possible.
If you have comments or remarks, please reach out to us.

%
%
%
\bibliographystyle{plainnat}

\bibliography{references}

\end{document}